\documentclass{pnastwo}

\usepackage{graphicx}
\usepackage{amssymb,amsfonts,amsmath}

\usepackage{multirow}
\usepackage{rotating}
\usepackage{cite}

\usepackage{color}

\begin{document}

\title{A self-learning algorithm for biased molecular dynamics}

\author{\normalsize Gareth A. Tribello,  Michele Ceriotti and Michele Parrinello\affil{1}{Computational Science, Department of Chemistry and Applied Biosciences, ETHZ Zurich USI-Campus, Via Giuseppe Buffi 13 C-6900 Lugano} }


             
\maketitle         
\newcommand{\mm}[1]{{\color{red}#1}}
\newcommand{\gt}[1]{{\color{blue}#1}}

\begin{article}
\begin{abstract} 
A new self-learning algorithm for accelerated dynamics, reconnaissance metadynamics, is proposed that is able to work with a very large number of collective coordinates.  Acceleration of the dynamics is achieved by constructing a bias potential in terms of a patchwork of one-dimensional, locally valid collective coordinates.  These collective coordinates are obtained from trajectory analyses so that they adapt to any new features encountered during the simulation.  We show how this methodology can be used to enhance sampling in real chemical systems citing examples both from the physics of clusters and from the biological sciences.
 \end{abstract} 
 

\section{Introduction}

\dropcap{M}any chemical systems, notably those in condensed matter and in biology, are characterized by the presence of multiple low energy states which are separated by large barriers.  The presence of these barriers prevents exploration of all of configuration space during the relatively-short timescales accessible in molecular dynamics (MD) simulations. Typically this means that only those configurations in a small, locally-ergodic region in the vicinity of the input structure are visited.  

A large number of methods have been put forward to overcome this difficulty, many of which either use some form of enhanced sampling \cite{PT1,PT2,sim-temp,solute-temp,multcanon-MD,wang-landau} or focus on the transition from one local minimum to another \cite{tps2,milestoning,rare-switching,string-method2}.  Other methods recognize that a small number of degrees of freedom (collective variables) accurately describe the interesting transitions and so either raise the temperature of these degrees of freedom \cite{TAMD1, TAMD2,TAMD4} or introduce a bias to enhance the sampling along them \cite{pmf,thermo_peturb,WH2,WH1,local_elevation,WH3,hyperdyn2,FE_from_average_force,self-healing-umbrella}. The major differences between the various approaches in this last class is the way in which the bias is generated a particularly useful technique being to use a bias that is dependent on the previously visited trajectory.  This is the basis of the metadynamics method \cite{metadyn,plumed} that has been introduced by our group and applied to a large variety of chemical problems \cite{metad-review}. 

For many chemical systems the interesting, reactive processes take place in a relatively low dimensional space \cite{piana-folding,clementi-low-dim-reduction}.  However, it is often not immediately obvious how to identify a set of collective variables (CVs) that span this 'reactive' sub-space.  Furthermore, in methods like metadynamics the presence of barriers in the transverse degrees of freedom leads to incomplete sampling.  The obvious solution therefore is to use very large numbers of collective coordinates.  However, although this is theoretically possible with methods such as metadynamics, it is impractical because the volume of bias that one must add to the free energy surface, and hence the length of the simulation, increases exponentially with dimensionality. One suggestion for resolving this issue is to run multiple short, 1D metadynamics simulations in parallel with different collective coordinates and to allow swaps between the different realizations based on a Monte Carlo criterion \cite{bias-exchange-metad}.  Here we propose an alternative solution based on the realization that, if the free energy surface is to be flattened, the majority of the bias will have to be added at or near the minima in the surface.  Identifying the locations of these minima is straightforward as, during a dynamical trajectory, the system should spend the majority of its time trapped in the vicinity of one or more of them \cite{gamus}.  Therefore by using a form of ``smart" bias that targets these low free energy regions specifically we can force the system away from them and into unexplored areas of configuration space.  We call the self learning algorithm that we have developed based on these ideas Reconnaissance Metadynamics and have implemented it in the plugin for molecular dynamics PLUMED \cite{plumed}.  In what follows we demonstrate this algorithm on two different systems - a cluster of seven Lennard Jones atoms and a short protein.  

\section{Background}

Before introducing our new method a brief survey of established techniques for dealing with complex energy surfaces in terms of very large numbers of collective coordinates is in order.  Zhu \emph{et al} \cite{zhu_tuckerman,minary_tuckerman} have introduced a method that uses a variable transformation to reduce barriers and thereby increase sampling, which works with a large number of collective coordinates.  Problematically though this method does not work if there are hydrogen bonds present.  In contrast, methods that work by thermostating the CVs at a higher temperature \cite{TAMD1, TAMD2,TAMD4} suffer no such problems and have been used successfully with large numbers of CVs \cite{TAMD3}.  However in these methods, unlike metadynamics, there is nothing that prevents the system from revisiting configurations, which could prove problematic for examinations of glassy landscapes with many local minima of equal likelihood.      

For many free energy methods explicitly including a large numbers of collective coordinates is not feasible.  However, one can use collective coordinates that describe a collective motion that involves many degrees of freedom.  For example, one can use the principle components of the covariance matrix of a large set of collective coordinates, the values of which have been calculated over a short MD trajectory \cite{essential-dynamics,conf-flooding}.  Alternatively, one can take non-linear combinations by defining a path in the high-dimensionality CV space \cite{A-B}.  The distance along and the distance from this path span a low-dimensional, non-linear space and metadynamics simulations using these two CVs have been shown to work well.  However, a great deal of insight is required in choosing an initial path from which the CVs are generated as the simulation can only provide meaningful insight for the region of configuration space in the immediate vicinity of this path.    

Entropy plays an important role in free energy surfaces as it can wash out potential energy minima and make it such that finite temperature equilibrium states do not correspond to minima in the potential energy surface.  Nevertheless, for systems with deep minima in the potential energy surface, one can use algorithms that locate all the minima on the 0~K (potential) energy surface and assume that the properties at every point on this surface are the same as those of the nearest local minima safe in the knowledge that the entropic effects are small.  These algorithms allow one to divide up configuration space and map every point to its appropriate local minimum using a minimization algorithm \cite{energy-landscapes}.  Furthermore, the slow modes in the vicinity of each basin provide good local collective coordinates as in the vicinity of basins the system is very nearly harmonic.  From a patchwork of such descriptors one could conceive of ways to obtain globally-non-linear CVs.  Recently, Kushima \emph{et. al.} \cite{kushima-supercooled} have developed a self-learning, metadynamics-based algorithm based on these ideas that works by minimizing the energy and adding bias functions at the position of the minimum found so that subsequent minimizations will identify new minima.       

Ideas described above for the exploration of potential energy surfaces cannot be straightforwardly transferred to the study of free energy surface because of the difficulties associated with the calculation of derivatives at finite temperature.  Nonetheless, Maragakis \emph{et al} \cite{gamus} have developed a method, GAMUS, that has some similarities to the 0~K method developed by Kushima et al.  In GAMUS the kinetic traps that are preventing free diffusion are located by fitting the probability distribution of visited configurations with a Gaussian Mixture (GM) model.  The resulting set of bespoke Gaussians are then used to update an adaptive bias that encourages the system to visit unexplored regions of configuration space.  This adaptive approach accelerates the filling of basins and thus provides a considerable speed up over conventional metadynamics when one is using 3 or 4 collective variables.  Nevertheless, the filling time will still increase exponentially with the number of CVs.    

\section{Reconnaissance metadynamics algorithm}
\label{sec:recon-metad} 

\begin{figure}
\includegraphics{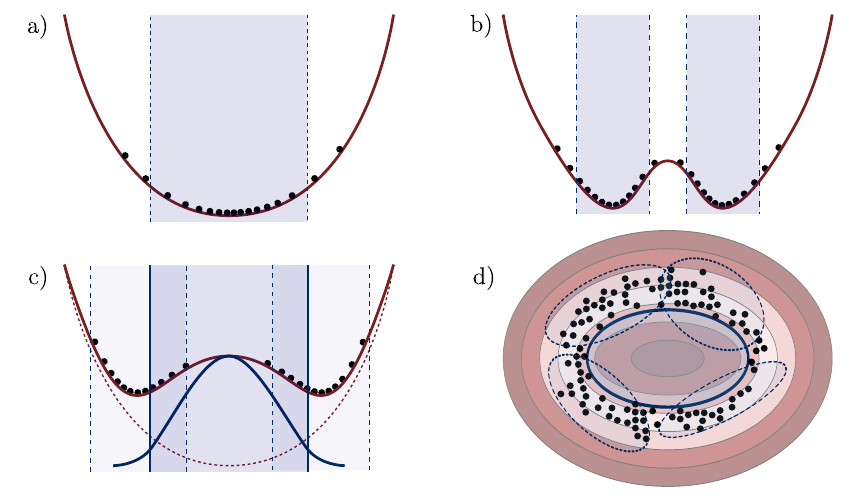}
\caption{Schematic representations of why it is necessary to use more than one pca analyzer and why it is necessary to expand basins.  The trajectory in panel (a) can be fit using a single pca analyzer.  By contrast the trajectory in panel (b) must be fit with a pair of analyzers as the energetic barrier is low enough that there will be hopping events between the two sub-basins.  Panel (c) shows how adding a single Gaussian to the center of the basin in panel (a) can lead to the creation of spurious basins in later cluster analyses.   Panel (d) demonstrates how this problem becomes more severe as the dimensionality is increased and also that it will occur if the basins are given a fixed size.}
\label{fig:expandbasfig}
\end{figure}

Reconnaissance metadynamics combines a number of new ideas with those of established methods and is thus effective with very large numbers of CVs.  The bias potential is constructed in terms of a patchwork of basins each of which corresponds to a low free energy region in the underlying FES.  These features/basins are recognized dynamically by periodically analyzing the trajectory with a sophisticated clustering strategy.  The region of configuration space in the vicinity of each basin is then described using a one-dimensional CV that is tuned using information collected during the clustering.  Consequentially, even when the overall number of collective coordinates ($d$) is large, depressions are compensated for rapidly because the bias is added in a locally-valid, low-dimensional space.    

\begin{figure}[h]
\includegraphics{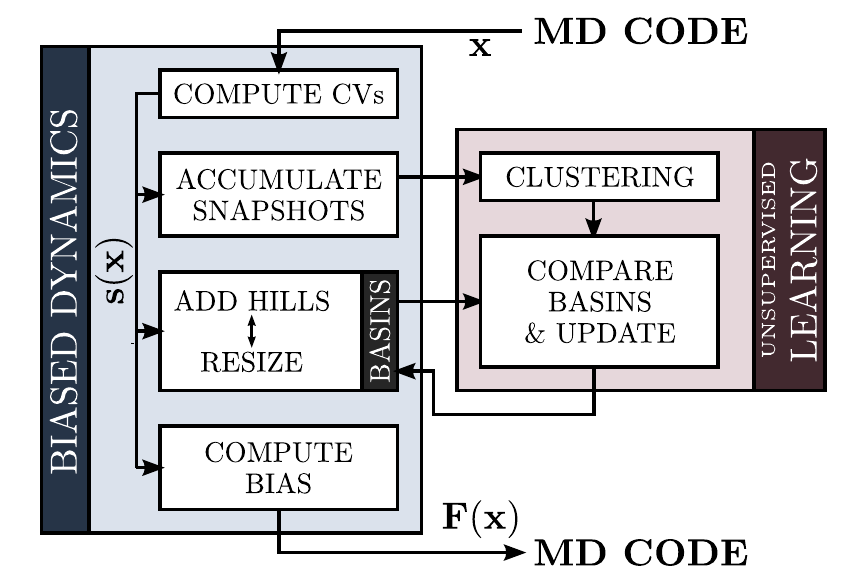}
\caption{Flow chart for the reconnaissance metadynamics algorithm.}
\label{fig:flow-chart}
\end{figure} 

Cluster analyses are performed at regular intervals using a set of stored configurations for the CVs that are accumulated from the trajectory.  During this analysis it is essential that some form of dimensionality reduction be performed as otherwise the fitting will be intractable.  In addition, one must recognize that, because this is a dynamical trajectory, the system may well have hopped between different basins on the free energy surface (see figure \ref{fig:expandbasfig}).   Therefore, because we would ideally like to treat each basin separately, principle component analysis (PCA) is not an option.  Furthermore, Gaussian mixture expectation maximization algorithms \cite{gamus},  although able to separate the basins, will become unstable when the number of collective coordinates is large.  Thankfully however combinations of these two algorithms exist \cite{PPCA,mix_ppca,ppca-anneal} that allow us to cluster the data while simultaneously reducing the dimensionality.  
 
This clustering strategy provides us with a set of Gaussian centers ($\boldsymbol{\mu}$) and covariance matrices ($\mathbf{C}$) for the various basins in the free energy surface.  Some of these will have very low weights or will be very similar to previously encountered basins and can thus be safely discarded.  Those remaining provide useful information on the local topology of the FES but cannot be used to predict the actual depth of the basin or its shape away from the center.  Consequentially, a flexible biassing strategy must be used in the vicinity of the minimum as addition of a single Gaussian will not necessarily compensate for the depression in free energy.  This failure to compensate basins fully can lead to the formation of spurious low energy features in the region surrounding the basin center (see figure \ref{fig:expandbasfig}(c)), which is a problem that becomes more severe as the dimensionality is increased.  To resolve these issues we assume that the basin is spherically symmetric in the metric induced by $\mathbf{C}$ and, in the spirit of metadynamics, construct an adaptive bias composed of small Gaussian hills of height $w_i$ and width $\Delta r$, along a single, radial collective coordinate $r(\mathbf{s})$ (see equation \ref{eqn:hills}).
%
\begin{eqnarray}
V_i(\mathbf{s}) &=& w_i  e^{ -\frac{ (r( \mathbf{s}) - r_{i})^2}{ 2\Delta r^2 } }  \label{eqn:hills} \\
r(\mathbf{s})^2 &=&  (\mathbf{s} - \boldsymbol{\mu} )^T \mathbf{C}^{-1} (\mathbf{s} - \boldsymbol{\mu} ) \label{eqn:non}
\end{eqnarray} 
In the above $\mathbf{s}$ is a vector that denotes the position in the full, $d$-dimensional CV space.  The form of equation \ref{eqn:hills} means that the bias associated with each hill acts in a hyper-ellipsoidal-crust-shaped region in the full, $d$-dimensional CV space.  Each of these crusts   has a shape much like a layer in an onion and so the integral of the bias added increases with $r(\mathbf{s})$.  Consequentially, the filling time for each basin no longer depends exponentially on the number of collective coordinates.  Furthermore, the hills have a shape that is consistent with the underlying anisotropy of the free energy surface because of the use of the covariance matrix in Eq.~\ref{eqn:non}.

Obviously the distance from a basin center is only a good collective coordinate when we are close to that center and so each basin must have a size, $S$.  This size assigns the region of configurational space in which it is reasonable to add hills that will force this system away from a particular basin.  Setting an initial value for this size is straightforward as we know from our fitting that the basin's shape is well described by a multivariate Gaussian.  Therefore, we choose an initial size equal to $S_{0} =  \sqrt{d -1} + 3$ because, as shown in the supplementary information, when the angular dependency of a $d$-dimensional Gaussian is integrated out the resulting distribution of $r$ is approximately equal to a 1D-Gaussian with standard deviation $\sqrt{2}$ centered at $\sqrt{d-1}$.            

If the size of basins is fixed then the problems described in figure \ref{fig:expandbasfig} are encountered once more.  Hence, in reconnaissance metadynamics the basin's size is allowed to expand during the course of the simulation so as to ensure that spurious minima, that would otherwise appear at the edge of basins, are dealt with automatically.  Periodically we check whether or not the system is inside the hyper-crust at a basin's rim ($S<r(\mathbf{s})<S+\Delta r$) and also that it is not within the sphere of influence of any other basin.  If these conditions are satisfied we then decide whether or not to expand using a probabilistic criterion, which ensures that it becomes more difficult to expand basins as the simulation progresses.  Based on the loose analogy with free particle diffusion outlined in the supporting information, this probability for expansion is given by $P=\textrm{min}(1,\frac{D\Delta t}{2\Delta r S})$, where $S$ is the current size of the basin, $\Delta t$ is the time between our checks on whether or not to expand and $D$ is a user defined parameter.  
  
The algorithm is summarized in the flow chart in figure \ref{fig:flow-chart}.  Further details on the components of the algorithm can be found in the methods section and in the supplementary information.   

\section{Results}

\subsection{2D-surface}
\label{sec:2d-surf}

\begin{figure}
\centering
\includegraphics{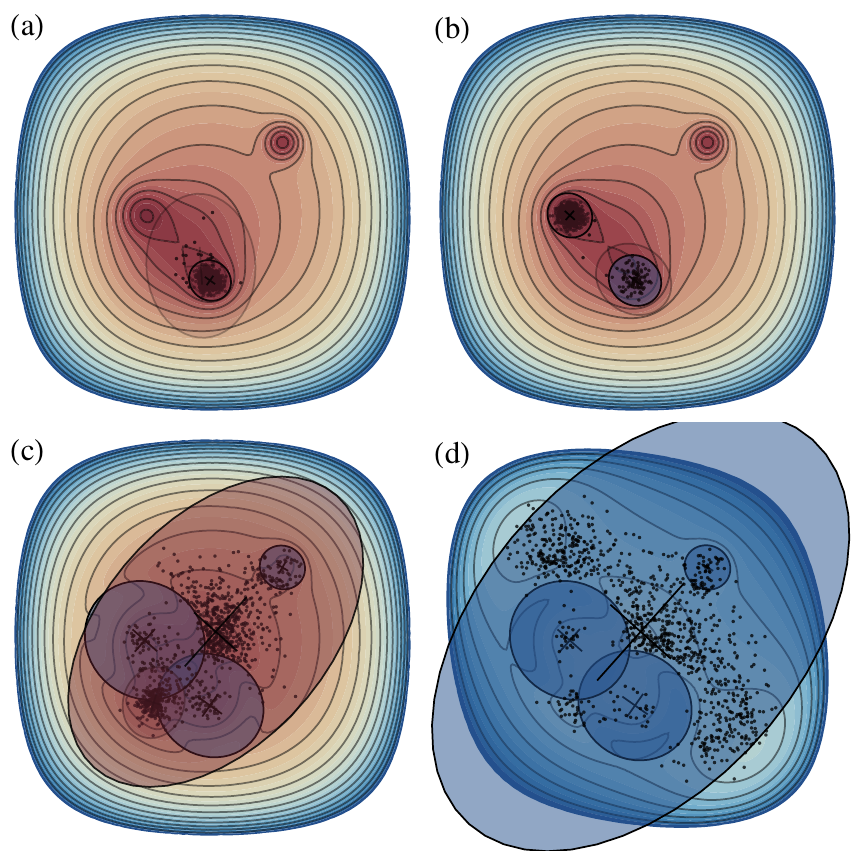}
\caption{Contour plots showing the potential energy + the current bias at selected points along a Reconnaissance Metadynamics trajectory for a particle diffusing about a 2D potential energy surface.  The black dots indicate the positions of the snapshots accumulated from the trajectory while the red ellipses indicate the basins found using the PPCA algorithm.  Blue ellipses are those basins, found during previous PPCA analyses, to which hills are being added.  The expansion of these blue basins as the bias grows is clearly seen in this figure.}         
\label{fig:2Dsurf}
\end{figure}

To illustrate the operation of the algorithm we first show how it can be used to accelerate the (Langevin) dynamics on the model, 2D potential energy surface illustrated in figure \ref{fig:2Dsurf}.  

At low temperatures a particle rolling about on the surface shown in figure \ref{fig:2Dsurf} will remain trapped in one of the deep basins.  This is precisely what is observed during the first part of the simulation, when no metadynamics is performed, as the first application of our clustering algorithm demonstrates.  On addition of bias, the system quickly escapes this first basin and falls into other basins, the locations of which are identified during subsequent applications of our unsupervised learning protocol.  This process continues throughout the simulation so, once basins are identified, the history dependent bias compensates for them quickly and hence the system rapidly explores the entirety of the energy surface.       

Figure \ref{fig:2Dsurf} shows that the reconnaissance metadynamics algorithm, when properly applied, finds only basins that correspond to the true features in the free energy surface.  In addition figure~\ref{fig:2Dsurf}(d) shows how effective the adaptive bias is in dealing with regions where small basins are encompassed in larger depressions. It clearly shows that initially small hills are used to deal with the sub-basins, while later much larger hills are used to compensate for the super-basin.  

\subsection{Lennard-Jones 7}
\label{sec:lj7}

\begin{figure*}
\centering
\includegraphics{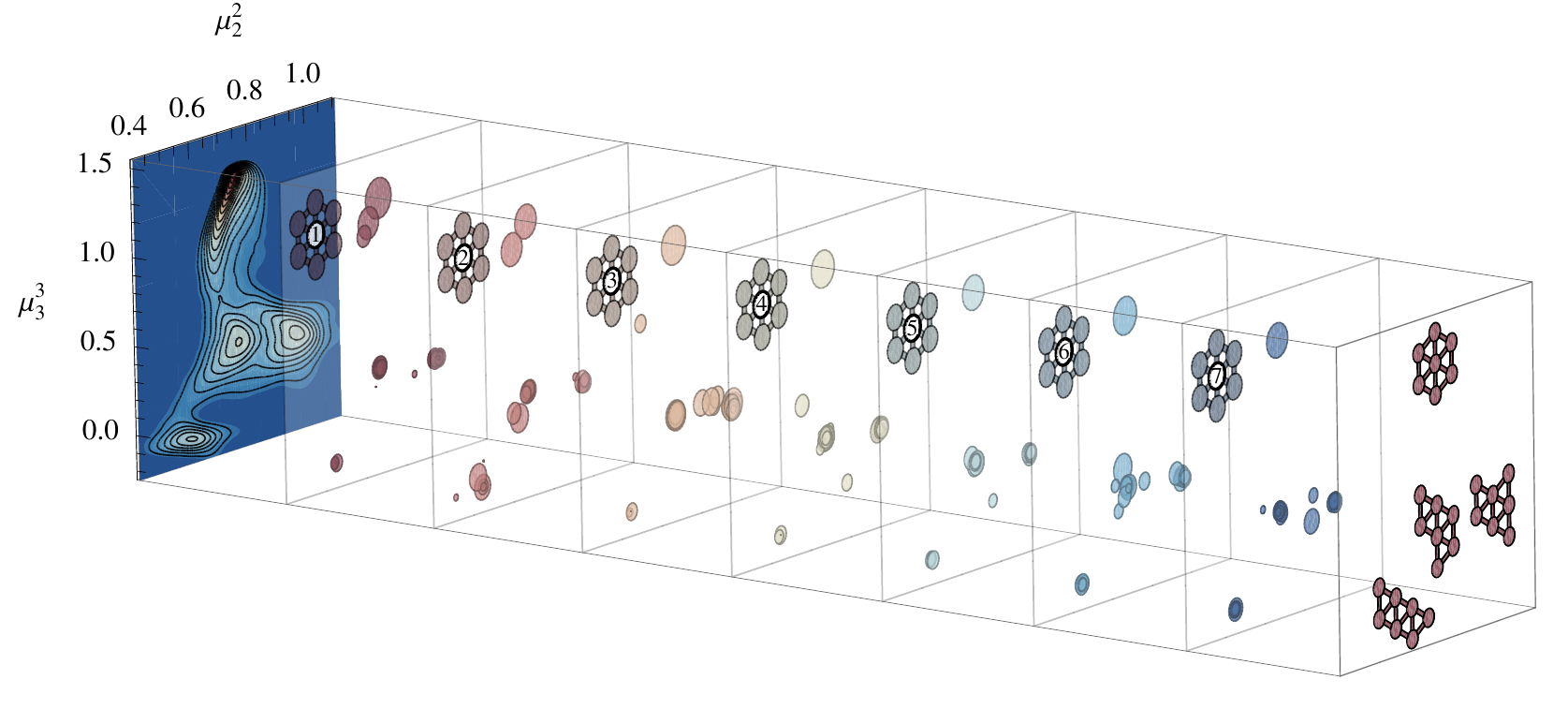}
\caption{The locations of the various basins found during a reconnaissance metadynamics simulation of the Lennard-Jones 7 cluster plotted as a function of the second and third moments of the distribution of coordination numbers and the index of the atom with the highest coordination number.  Also shown is a free energy surface calculated using a well-tempered metadynamics simulation that employed the second and third moments of the distribution as collective coordinates.  Each basin is represented by a circle with an area that is proportional to the bias at its center.  On the free energy surface contours are placed at intervals of 0.1~$\epsilon$}
\label{fig:LJfigure}
\end{figure*}

Small clusters of rare-gas atoms have a remarkably complex behavior despite their rather limited number of degrees of freedom.  A particularly well studied example is the two-dimensional, seven-atom, Lennard-Jones cluster \cite{tps1,vergillius} for which 4 minima and 19 saddle points in the potential energy surface have been identified \cite{lj-minima} (see figure \ref{fig:LJfigure}).  At moderate temperatures ($k_B T=0.1\epsilon$) this system spends the majority of its time oscillating around the minimum energy structure, in which one of the atoms is surrounded symmetrically by the six other atoms.  Infrequently however the system will also undergo isomerizations in which the central atom of the hexagon is exchanged with one of the atoms on the surface \cite{lj-minima}.  

To test the reconnaissance metadynamics algorithm we examined this system using the coordination numbers of all the atoms (seven collective coordinates).  In doing this we neglect the interchange symmetry to see if we can reproduce this symmetry in the positions of the various basins we find.  Figure \ref{fig:LJfigure} gives the results of this calculations and denotes the position of each of the basins found by a circle whose area reflects the final bias at the basin center.  The three lowest lying minima for this cluster have one atom which has a much higher coordination number than the others and so the basins identified along the trajectory have been grouped, based on the index of the atom with the highest coordination number, onto seven slices.  On each of these planes the positions and sizes of the basins are consistent, which suggests that the algorithm finds the correct symmetry and explores configuration space correctly. 

In reconnaissance metadynamics there is no straightforward connection between the bias and the free energy because each CV is only valid in a local region and hence the overall bias is not a function of a global order parameter.  Nonetheless, the bias greatly enhances the exploration of phase space and so free energies could be obtained using an umbrella sampling approach.  However, even without this additional step, a reconnaissance metadynamics trajectory gives one a feel for the lie of the land, which can be used to obtain chemical insight.  For example, the basin centers provide a set of landmark points that can be used to validate a low dimensionality description of the system \cite{clementi-low-dim-reduction}.  For this simple case we did this by calculating the value at the basin centers of many different candidate coordinates.  We discovered that an optimal, in-plane separation of the various basins is attained when we use the second and third moments ( $\mu_2^2=\frac{1}{N} \sum_{i=1}^{N} (c_i - \langle c \rangle)^2$ and $\mu_3^3=\frac{1}{N}\sum_{i=1}^{N} (c_i - \langle c \rangle)^3$ respectively)  of the distribution of coordination numbers.  These two CVs clearly project out permutation symmetry and so we also performed a conventional well-tempered metadynamics simulation.  Figure \ref{fig:LJfigure} shows the free energy energy surface obtained in its eighth slice.  A comparison of this surface with the results from the reconnaissance metadynamics shows how the basins found cluster around the minima in this FES.        

\subsection{Poly-alanine 12}
\label{sec:protein}

The protein folding problem is commonly tackled using computer simulation and there exist model systems for which the entirety of the potential energy landscape has been mapped out \cite{energy-landscapes} that represent a superb test of any new methodology.  For example polyalanine-12, modeled with a distance dependent dielectric ($\epsilon_{ij}=r_{ij}$ in Angstroms) that mimics some of the solvent effects, has been extensively studied  \cite{wales-ala12}.  This protein has a funnel-shaped, energy landscape with a alpha-helical, global minima.  We found that during a 1~$\mu$s, conventional MD simulation started from a random configuration, the protein did not fold (see supplementary information).  Hence, examining whether or not the protein will fold during a reconnaissance metadynamics simulation will provide a third test of our methodology. 

\begin{figure*}
\centering
\includegraphics{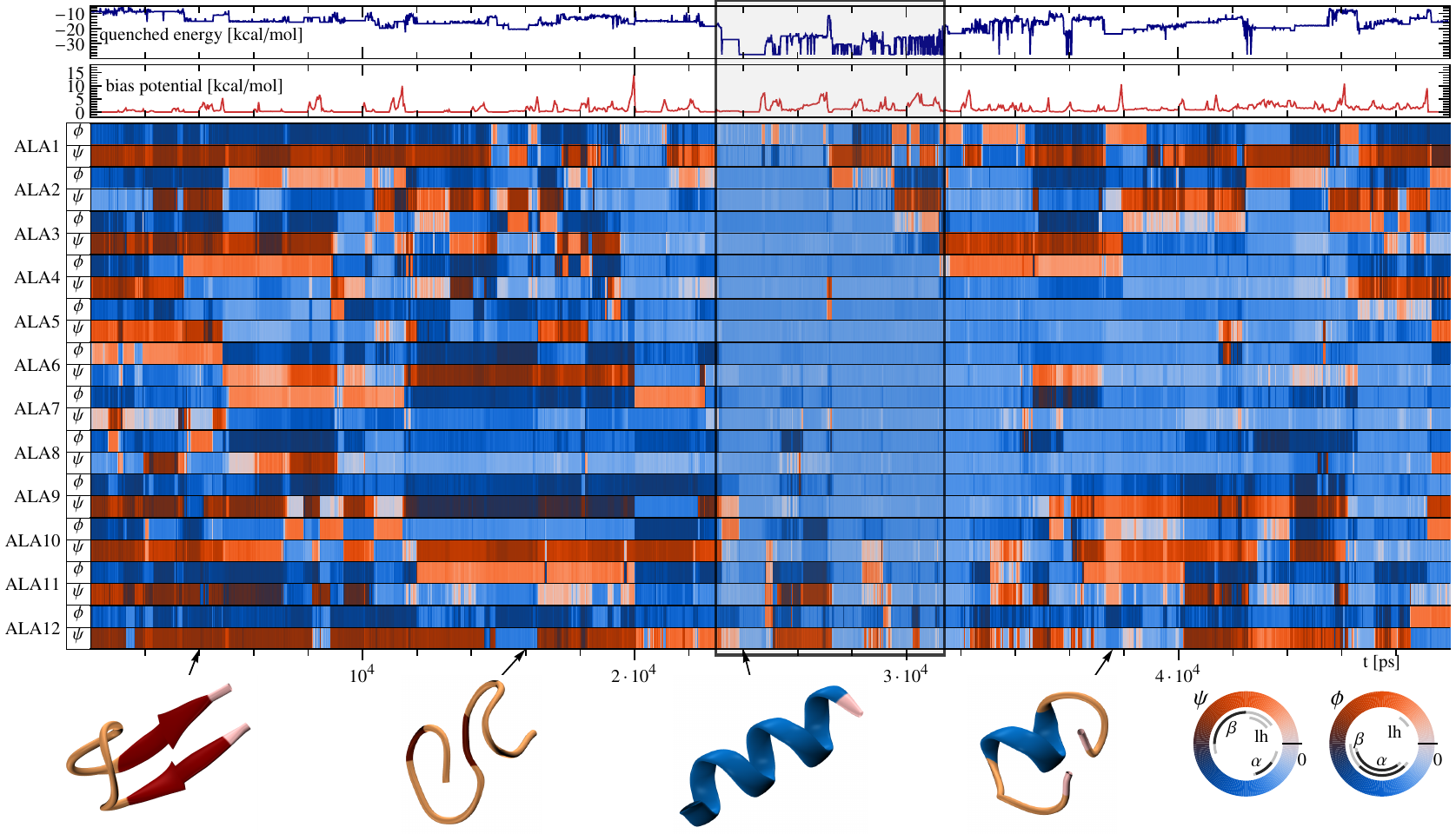}
\caption{A representation of a 50 ns portion of the reconnaissance metadynamics trajectory for alanine 12 modeled with a distance dependent dielectric.  The bars in the lower panel give the values of 40~ps running averages for each of the torsional angles in the protein (see key).  The red (light) line shows a similar running average for the bias potential.  The system was annealed every 20~ps to examine the transitions between inherent structures and the energies obtained are indicated by the blue (dark) line. A number of low-energy, representative structures found during the trajectory are also shown and the box highlights the portion of the trajectory when the protein has a configuration that is near the global-minimum, alpha-helix structure.  }
\label{fig:ala12}
\end{figure*}

For this reconnaissance metadynamics calculation we used the 24 backbone dihedral angles as the collective coordinates as these angles provide an excellent description of the protein structure.  These variables are periodic, which had to be accounted for in the method by replacing the multivariate Gaussians with multivariate von Mises distributions \cite{von-misses}.  This distribution, if sufficiently concentrated about the mean, is equivalent to a Gaussian in which the difference between any point and the mean is shifted to the minimum image.  Consequentially, we can continue to use the same algorithm for trajectory analysis as long as we take into account the periodicity when we calculate differences and averages.  In addition, we can define a quantity (see equation \ref{eqn:per}) that is equivalent to the distance from the center of the basin (equation \ref{eqn:non}) but that takes into account the periodicity of the CVs ($P_i$).

\begin{eqnarray}
r(\mathbf{s})^2 &=& 2 \sum_{i=1}^{d} C^{-1}_{ii}\left[1-\cos\left( \frac{2\pi [s_i-\mu_i]}{P_i}\right) \right] \label{eqn:per} \\
&+& \sum_{i \ne j} C^{-1}_{ij} \sin\left( \frac{2\pi [s_i-\mu_i]}{P_i}\right) \sin\left( \frac{2\pi [s_j-\mu_j]}{P_j}\right)   \nonumber
\end{eqnarray}

Figure~\ref{fig:ala12} provides a representation of a portion of a typical reconnaissance metadynamics trajectory of the protein.  The figure shows the values of all the backbone torsional angles and clearly demonstrates that a large volume of configurational space is explored during this relatively-short simulation.  Furthermore, unlike in conventional MD, after approximately 22~ns the protein folds into the global-minimum, alpha-helix configuration.  To further demonstrate that the reconnaissance metadynamics is ensuring that large portions of configurational space, that would otherwise not be visited, are being explored we calculated the inherent structures by minimizing the energy every 20~ps.  The energies of the minimized structures are shown in the uppermost panel of figure~\ref{fig:ala12} and demonstrate that the potential energy surface is very rough and that the alpha helix is considerably lower in energy than the other energy minima encountered.            

\section{Conclusions}

We have proposed a new accelerated dynamics scheme, reconnaissance metadynamics, that uses a self-learning algorithm to construct a bias that accelerates the exploration of configuration space.   This algorithm works by automatically identifying low free energy features and deploying a bias that efficiently and rapidly compensates for them.  The great advantage of this algorithm is that there is no limit on the number of CVs on which the bias acts.  This is only possible because we collapse all these CVs into a single collective coordinate that is only valid locally and patch together a number of these local descriptors in order to reflect the fact that the important degrees of freedom are not uniform throughout configuration space.  As shown in figure \ref{fig:2Dsurf} this approach provides us with a simple, compact, hierarchical description of the free energy surface, which could be used to construct bias potentials for umbrella sampling.   

As shown for the two chemical systems on which we have demonstrated the algorithm, our ability to work with large numbers of collective coordinates means that one can employ generic configurational data, such as torsional angles and coordination numbers, as collective coordinates and thereby avoid all the usual difficulties associated with choosing a small set of collective coordinates.  In fact, even when CVs, on which there are no large barriers to motion, such as the torsional angles on the terminal amino acid groups in ala12, are included the algorithm will still function.  For both systems our method produces trajectories that contain an extensive exploration of the low energy parts of configurational space and hence provides a feel for the lie of the land.  Furthermore, even without a quantitative estimate of the free energy, considerable insight can be obtained from the trajectory.  In the Lennard Jones this allowed us to attain an effective two dimensional description of the landscape.  For more complex systems non-linear embedding could be used to automate this procedure.    

\begin{materials}

\section{Mixture of probabilistic principle component analyzers} Throughout this work we use the annealing strategy outlined in reference \cite{ppca-anneal} and in the supplementary information to do clustering.   This algorithm requires one to state at the outset the number of clusters that are being used to fit the data and the number of annealing steps.  For the latter we initially set $\scriptstyle{\sigma^2}$, which is the quantity treated like the temperature during the annealing, equal to the maximum eigenvalue of the covariance of the data and lowered $\scriptstyle{\sigma^2}$ until it was less than 1 \% of its initial value.  To establish the correct number of clusters we run multiple fits to the data using different numbers of clusters and select the fit that gives the largest value for the Bayesian Information Criterion \cite{BIC}, $\scriptstyle{\textrm{BIC} = 2 \log [ \mathcal{L}(\mathbf{x},\boldsymbol{\theta}) ] - n_p \log[M]}$, where $\scriptstyle{\mathcal{L}(\mathbf{x},\boldsymbol{\theta})}$ is the maximized likelihood for the model, $\scriptstyle{n_p}$ is the number of parameters in the model and $\scriptstyle{M}$ is the number of trajectory snapshots used in the fitting.  
 
\section{Selecting novel basins} As already discussed the GM algorithm provides us with the locations of a number of basins.  Some of these will have very low weights in the fit and can therefore be safely ignored in the construction of the bias.  Others however will provide information about the basins found during prior runs - in short information that is redundant.  Therefore, we introduce a criterion for the selection of basins that requires that $\scriptstyle{f_i[1-\textrm{max}(\xi_{ij})] > \textrm{TOL}}$, where TOL is some user defined tolerance, $\scriptstyle{f_i}$ is the weight of the new basin in the fit and $\scriptstyle{\xi_{ij}}$ is the similarity between the new basin $\scriptstyle{i}$ and the old basin $\scriptstyle{j}$.  To calculate $\scriptstyle{\xi_{ij}}$ we use Matusita's measure which can be calculated exactly for multivariate Gaussians \cite{mautista}.  For an expanded basin of size $S$ its original covariance is multiplied by a factor of $\scriptstyle{S/S_{0}}$ when calculating this function so that their expanded volume is appropriately taken into account.  

\section{Lennard-Jones} The parameters for the simulations of Lennard-Jones 7, in Lennard Jones units are as follows.  The temperature was set equal to ${\scriptstyle {k_BT}=0.1{\epsilon}}$ using a Langevin thermostat, with a relaxation time of 0.1~${\scriptstyle \sqrt{\epsilon/m \sigma^2}}$.  The equations of motion were integrated using the velocity verlet algorithm with a timestep of 0.01~${\scriptstyle \sqrt{\epsilon/m \sigma^2}}$ for 5~$\scriptstyle{\times}$~10$^7$ steps.  During reconnaissance metadynamics the CVs were stored every 100 steps, while cluster analysis was done every 1~$\scriptstyle{\times}$~10$^5$ steps.  Only basins with a weight greater than 0.3 were considered and to these attempts to add hills of height 0.5~$\scriptstyle{k_B T}$ and a width of 1.5 were made every 1000 steps.  Basin expansion was attempted with the same frequency with the parameter $\scriptstyle{D}$ set equal to 0.03.  The coordination numbers were computed using  ${\scriptstyle c_i =\sum_{i \ne j}{1-\left(\frac{r_{ij}}{1.5} \right)^8}\left[1-\left(\frac{r_{ij}}{1.5}\right)^{16} \right]^{-1}}$, where ${\scriptstyle r_{ij}}$ is the distance between atoms ${\scriptstyle i}$ and ${\scriptstyle j}$. 

\section{poly-alanine-12} All simulations of polyalanine were run with a modified version of gromacs-4.0.3 \cite{gromacs}, the amber96 forcefield \cite{amber96} and a distance dependent dielectric.  A timestep of 2~fs was used, all bonds were kept rigid using the LINCS algorithm and the van der Waals and electrostatic interactions were calculated without any cutoff.  The global thermostat of Bussi et al \cite{bussi_therm} was used to maintain the system at a temperature of 300~K.  The initial random configuration of the protein was generated by setting up the protein in a linear geometry, minimizing it and then running 1~ns of normal MD at 300~K in order to equilibrate.     During reconnaissance metadynamics the CVs were stored every 250 steps, while cluster analysis was done every 5~$\scriptstyle{\times}$~10$^5$ steps.  Only basins with a weight greater than 0.2  were considered and attempts were made every 1000 steps to add to these basins hills of height 0.4~$\scriptstyle{k_B T}$ and width 1.5.  Basin expansion was attempted with the same frequency with the parameter $\scriptstyle{D}$ set equal to 0.3.  Minimizations to obtain inherent structures was done by first annealing for 1.2~ns with a decay rate of 0.996~ps$^{-1}$ and subsequently performing a steepest decent minimization.  Guidelines as to how to select parameters for reconnaissance metadynamics are provided in the supplementary information. For both this system and the Lennard Jones cluster we obtained similar results with a variety of different parameters sets.  

\end{materials}   

\begin{acknowledgments}
The authors would like to thank Davide Branduardi, Jim Pfaendtner, Meher Prakash and Massimilliano Bonomi for useful discussions.  David Wales is also thanked for his advice on the simulation of the ala12 system.   
\end{acknowledgments}

\end{article}

\end{document}


\title{SUPPLEMENTARY INFORMATION A self-learning algorithm for biased molecular dynamics}

\author{\normalsize Gareth A. Tribello,  Michele Cerriotti and Michele Parrinello }


             
\maketitle    


In the following the mathematical details for the trajectory analysis tools used in the reconnaissance metadynamics algorithm are provided along with some information on how to select parameters.  In addition, we also provide a collection of the low energy structures we found for ala12 and an analysis of a conventional MD trajectory for this system.

\section{Mathematical details}

As discussed in the paper the algorithm we use for clustering is taken from references \cite{ppca-resolution,ppca-anneal}.  In the following some background information on this algorithm is provided, as some of the concepts behind it are perhaps unfamiliar to those working in the molecular simulation community.

\subsection{Expectation-Maximization}

An expectation maximization algorithm is a method for fitting data to a statistical model.  To understand it consider a collection of m points, $\mathbf{s}_m$  - these could come from a dynamical trajectory.  One can easily calculate the probability that these points came from a particular statistical distribution, $P(\mathbf{s}_m)$ by computing the following:
\begin{equation}
\mathcal{L}  = \prod_{m=1}^M P(\mathbf{s}_m)
\end{equation}   
This quantity is called the likelihood of the model and is often quoted in terms of its logarithm as typically it will be very small.  If instead of having a fixed probability model one has a model that depends on some set of parameters, or latent variables $\boldsymbol{\theta}$, then the likelihood will clearly be a function of these parameters:
\begin{equation}
\mathcal{L}(\boldsymbol{\theta})  = \prod_{m=1}^M P(\mathbf{s}_m,\boldsymbol{\theta})
\end{equation}       

Obviously, when fitting the parameters of the model one would like to maximize this likelihood.  Commonly this is done using an expectation maximization algorithm.  This iterative algorithm works by performing alternate expectation and maximization cycles.  In the expectation (E) step the values for the probabilities of all the points are calculated from the current estimate of the probabilistic model.  These probabilities can then be used to compute the likelihood of the model.  In addition the calculated probabilities are used in the maximization step to recalculate the parameters of the model. Re-performing the E-step with the new estimate of the model parameters gives an improved likelihood and hence this process of alternate E and M steps eventually brings one to a maximum-likelihood solution.   

\subsection{Gaussian mixture models}
\label{sec:GM}

The Gaussian mixture (GM) model \cite{numerical-recipes} is one of the simplest examples of classification by unsupervised learning and is the algorithm used in the GAMUS methodology \cite{gamus}.  This algorithm aims to fit a probability distribution composed of $N$ multivariate Gaussian distributions in  $d$ dimensions using an input set of $M$ points.  This is achieved by fitting the centers $\boldsymbol{\mu}_n$, covariances $\mathbf{\Sigma}_n$ and weights $f_n$ of the Gaussian distributions using an expectation maximization algorithm.  During the E-step the probability for each point in each of the Gaussian distributions is calculated using:
\begin{equation}
 P_n( \mathbf{s}_m) =   \frac{1}{  (2\pi)^{\frac{d}{2}} | \boldsymbol{\Sigma_n} |^{\frac{1}{2}} }     \exp \left[ -\frac{1}{2} (\mathbf{s}_m - \boldsymbol{\mu}_n )^T \boldsymbol{\Sigma}_n^{-1}  (\mathbf{s}_m - \boldsymbol{\mu}_n )  \right] 
\end{equation} 
where $ | \boldsymbol{\Sigma_n} |$ is the determinant of  $\mathbf{\Sigma}_n$.  The likelihood of the model is then computed using:
\begin{eqnarray}
P(\mathbf{s}_m) &=& \sum_{n=1}^N f_n P_n(\mathbf{s}_m) \\
\mathcal{L} &=& \prod_{m=1}^{M} P(\mathbf{s}_m)
\end{eqnarray}
Having computed this probability data one can then perform the M-step in which new parameters for the probability model are calculated using:
\begin{eqnarray}
\Psi_n &=& \sum_{m=1}^{M} P_n( \mathbf{s}_m) \qquad \qquad \qquad \qquad f_n = \frac{\Psi_n}{M}  \label{eqn:weight} \\
\boldsymbol{\mu}_n &=& \frac{1}{\Psi_n} \sum_{m=1}^{M}  \mathbf{s}_m P_n( \mathbf{s}_m )   \label{eqn:mean} \\  
\boldsymbol{\Sigma}_n &=& \frac{1}{\Psi_n} \sum_{m=1}^{M} (\mathbf{s}_m - \boldsymbol{\mu}_n) (\mathbf{s}_m - \boldsymbol{\mu}_n)^T P_n( \mathbf{s}_m)   \label{eqn:covar}
\end{eqnarray}
Iterating these two steps leads one to a maximum in the likelihood function. \\

There are two major disadvantages of this algorithm.   The first is that the number of parameters that must be fitted increases quadratically with the number of dimensions.  Consequentially, it is only possible to use Gaussian mixture models with low dimensionality data.  The second major problem is that it is not obvious how to initialize the parameters during the expectation step.  This is important because there are many local maxima of the likelihood and therefore different start points will lead to different solutions.  A common way of sidestepping this issue is simply to run the algorithm multiple times and to pick the best fit of the data.

\subsection{Deterministic annealing}

Often, one's concern is hard clustering; namely, classifying a set of data points into a small number of clusters based on the spatial distribution.  Obviously, if this is the aim then one is not so concerned with fitting the data to a probability distribution using the methods described in the previous section.  Nevertheless hard-clustering algorithms work much like the EM algorithm described in the previous section \cite{numerical-recipes}.  Again though a problem with these algorithms is the presence of multiple local maxima in the likelihood function.  Hence, starting the optimization with different initial conditions will lead to different solutions.  

For these hard clustering problems Rose \emph{et al} \cite{DA_prl} have come up with an elegant solution to the initialization problem that is based on using an annealing strategy.  This method works by doing multiple fits of the data to a probability distribution that is composed of $N$ spherical Gaussian distributions.  At each stage they fit the weights and centers of the Gaussian distributions.  Meanwhile, the standard deviation, $\sigma$, of the distribution is treated like a temperature and is thus exponentially decreased as the annealing progresses.  For each value of $\sigma$ the fitting is done using an expectation maximization strategy in which, during the E-step, the probabilities of the data in each of the Gaussian distributions is computed using:
\begin{equation}
 P_n( \mathbf{s}_m) =   \frac{1}{  (2\pi \sigma)^{\frac{d}{2}} }  \exp \left[ -\frac{1}{2 \sigma^2} (\mathbf{s}_m - \boldsymbol{\mu}_n )^T  (\mathbf{s}_m - \boldsymbol{\mu}_n )  \right] 
\end{equation}
and the likelihood is computed using:
\begin{eqnarray}
P(\mathbf{s}_m) &=& \sum_{n=1}^N f_n P_n(\mathbf{s}_m) \\
\mathcal{L} &=& \prod_{m=1}^{M} P(\mathbf{s}_m)
\end{eqnarray}
Then during the M step the weights and centers of the distribution are re-computed using equations \ref{eqn:weight} and \ref{eqn:mean} respectively.

The advantage of the annealing is that it is less likely to get trapped at local maxima of the likelihood surface.  Furthermore, finding an initial start point is straightforward because, for high values of $\sigma$, all the clusters will coalesce into a single cluster.  Hence, the calculation can be initialized with a configuration in which all the cluster centers are placed at the center of mass of the data.  In fact Rose \emph{et al} \cite{DA_prl} have demonstrated, through an analogy with statistical mechanics, that as $\sigma$ (which is the analogue of temperature) is lowered the system will undergo a sequence of phase transitions.  During each of these phase transitions Gaussian distributions, which prior to the transition would merge on a single center, split apart in order to reflect the clustered structure inherent in the data.

\subsection{Probabilistic principle component analysis}

Principle component analysis (PCA), or essential dynamics, is a popular technique in the field of molecular simulation.  It is commonly used to obtain directions in which there is maximal change in some particularly interesting set of collective coordinates be they the coordinates of particular atoms \cite{essential-dynamics}, the dihedral angles or the distances between particular atoms or groups of atoms.  In PCA one computes the mean and covariance of the collective coordinates from a short MD trajectory.  One then diagonalizes this covariance and projects the molecular motions onto the eigenvectors with the highest eigenvalues.  This is useful as it gives one a sense for directions in which the structural changes are maximal.  However, it is important to realize that these directions of maximal structural change are only valid in the small, locally-ergodic region which was explored during the unbiassed MD trajectory \cite{pca_is_rubbish}.  In other words, in unexplored regions of configurational space there is no guarantee that the principle components, calculated using essential dynamics, are the directions in which the structural changes are maximal.  Consequentially, one would ideally like to fit the size of the region over which the calculated principle components are valid, which would obviously be related to the region of phase space explored during the course of the trajectory. 

In the machine learning literature PCA algorithms exist that are able to fit both the principle components and the region of space over which these components are valid \cite{PPCA}.  These so called probabilistic principle component analysis algorithms (PPCA) work by assuming that the data is distributed in $d$-dimensional space according to a multivariate Gaussian distribution:
\begin{equation}
P( \mathbf{s}_m) =   \frac{1}{  (2\pi)^{\frac{d}{2}} | \mathbf{C} |^{\frac{1}{2}} }     \exp \left[ -\frac{1}{2} (\mathbf{s}_m - \boldsymbol{\mu}_n )^T \mathbf{C}^{-1}  (\mathbf{s}_m - \boldsymbol{\mu}_n )  \right] 
\label{eqn:ppca}
\end{equation}  
Here the covariance, $\mathbf{C}$, is given by:
\begin{equation}
\mathbf{C} = \sigma^2 \mathbf{I} + \mathbf{W}\mathbf{W}^T
\end{equation}  
where $\mathbf{W}$ is a $(d\times q)$ parameter matrix that describes the distribution in the directions in which there are are large fluctuations.  The fluctuations in the remaining directions are assumed less important.  Hence, they are dealt with in an averaged way by assuming that in these directions the data can be modeled using a spherical Gaussian with a standard deviation of $\sigma$.  

The existence of this probability distribution allows one to define a likelihood for the data which is given by:
\begin{equation}
\log \mathcal{L} = \sum_{m=1}^{M} \log P(\mathbf{s}_m) = -\frac{M}{2}  \left[ d \ln (2\pi) + \ln | \mathbf{C} | + \textrm{Tr}( \mathbf{C}^{-1} \boldsymbol{\Sigma} ) \right]
\end{equation}
where $\boldsymbol{\Sigma}$ is the covariance of the $d$-dimensional data, which can be easily computed.  Stationary points for this likelihood can be computed and it emerges that the global maximum for the likelihood occurs when $\mathbf{W}$ is given by:
\begin{equation}
\mathbf{W} = \mathbf{U}_q( \mathbf{\Lambda}_q - \sigma^2 \mathbf{I} )^{\frac{1}{2}}
\end{equation}
where the $q$ column vectors in the $(d\times q)$ matrix $\mathbf{U}_q$ are the $q$ principle eigenvectors of the covariance and $\mathbf{\Lambda}_q$ is a diagonal matrix that contains the corresponding eigenvalues.  The maximum likelihood estimate of $\sigma^2$ meanwhile can be calculated as the average of the remaining eigenvalues.  Alternatively one can fix $\sigma^2$ at the outset and can set $q$, the number of principle components in $\mathbf{W}$, equal to the number of eigenvectors whose corresponding eigenvalues are greater than $\sigma^2$.  

\subsection{Mixtures of probabilistic principle component analyzers}

As discussed in the previous section in the probabilistic version of principle component analysis (PPCA) \cite{PPCA} there is a probabilistic model for the observed data.  What is more this probability distribution is a multivariate Gaussian.  This suggests that one can combine the PPCA framework with the GM model and fit the data as a mixture of probabilistic principle component analyzers \cite{mix_ppca} thereby reducing the number of parameters required for clustering in high dimensionality.  There are only two differences in the resulting algorithm from that described above.  The first is that during the E-step, the $\mathbf{C}$ matrix obtained from the principle components must be used in computing the probabilities of the points rather than the true covariance, $\mathbf{\Sigma}$, of the data:
\begin{equation}
P_n( \mathbf{s}_m) =   \frac{1}{  (2\pi)^{\frac{d}{2}} | \mathbf{C}_n |^{\frac{1}{2}} }     \exp \left[ -\frac{1}{2} (\mathbf{s}_m - \boldsymbol{\mu}_n )^T \mathbf{C}_n^{-1}  (\mathbf{s}_m - \boldsymbol{\mu}_n )  \right] 
\end{equation}  
The second difference is that after updating the weights, means and covariances of the Gaussians, using equations \ref{eqn:weight}, \ref{eqn:mean} and \ref{eqn:covar} respectively, one must diagonalize the covariance matrices so as to extract the principle eigenvalues and eigenvectors.  These can then be used to recompute $\mathbf{C}_n$:  
\begin{eqnarray}
\mathbf{W}_n &=& \mathbf{U}_{n,q} ( \mathbf{\Lambda}_{n,q} - \sigma^2 \mathbf{I})^{\frac{1}{2}}\mathbf{I} \\
\mathbf{C}_n &=& \sigma^2 \mathbf{I} + \mathbf{W}_n\mathbf{W}_n^T 
\end{eqnarray} 
where the $d$ x $q$ matrix $\mathbf{U}_{n,q}$ contains the $q$ principle components of the covariance matrix $\mathbf{\Sigma}_n$ and $\mathbf{\Lambda}_q$ is a diagonal matrix that contains the corresponding eigenvalues.

\subsection{A simulated annealing algorithm for PPCA clustering}

The algorithm used in this work for doing clustering \cite{ppca-resolution,ppca-anneal} combines all of the elements discussed in the previous sections.  Hence, we are using the mixture of PPCA analyzers discussed above but are using an annealing strategy to do the fitting rather than relying the EM algorithm alone.  
The resulting algorithm reads:
\begin{algorithmic}
\STATE Place all clusters at the center of mass of the data set.
\STATE Set $\sigma^2$ equal to maximum eigenvalue of covariance $\lambda_{\textrm{max}}$.
\WHILE{Negligible separation between two cluster centers}
\STATE {Perturb position of the center of all clusters. }
\STATE {Perform EM minimization with spherical covariance. }
\STATE {$\sigma\leftarrow \alpha \sigma$}
\ENDWHILE
\WHILE{$\sigma^2>0.01 \lambda_{\textrm{max}}$}
\STATE {Perform EM using equations 1-6 }
\STATE {$\sigma\leftarrow \alpha \sigma$}
\ENDWHILE
\end{algorithmic}
In the early phases one performs a deterministic annealing, which continues until all the clusters are well separated.  
Then, in the final stages of fitting, one uses the mixture of probabilistic principle component analyzers. 
For each E-M step, the number of principle components $q$ is set equal to the number of 
eigenvalues of the covariance matrix that are larger than the current value for $\sigma^2$.  

\section{Some useful properties of multivariate Gaussians}

As discussed we use some known properties of multivariate Gaussians in the algorithm.  Firstly, we use the fact that integrating out the angular distribution in this function gives the following dependence of the distribution on the distance from the 
mean in the metric of the covariance, $r$:
\begin{eqnarray}
S(r') &=& \int .. \int \exp \left[ -\frac{ \left( ( \mathbf{s} - \boldsymbol{\mu} )^T \mathbf{C}^{-1} (\mathbf{s} -\boldsymbol{\mu} ) \right)^2}{2} \right] \delta(r(\mathbf{s}) - r') \textrm{d}^d\mathbf{s} \nonumber \\
&=&  \int ..\int \exp \left( -\frac{r^2}{2} \right) r^{d-1}  \delta(r -r' ) \textrm{d}r \textrm{d}S \nonumber \\
&=& S(d, \mathbf{C} ) r'^{d-1} \exp \left( -\frac{r'^2}{2} \right) \nonumber \\
&\approx&  S(d, \mathbf{C} )  \exp \left( -(r' - \sqrt{d-1})^2  \right)
\end{eqnarray}
Here $S(d, \mathbf{C} )$ indicates the surface integral of a hypersphere of dimension $d$ transformed by a orthogonal matrix $\mathbf{C}$.  The result in  the final step, that the distribution as a function of $r$ is approximately a Gaussian centered at $\sqrt{d -1}$, can be proved by comparing term by term the series expansions about $\sqrt{d -1}$ for the two functions.   

Matusita's measure for two multivariate Gaussian  can be computed explicitly as:
\begin{eqnarray}
\xi_{ij} &=& \int \sqrt{P_i(\mathbf{x})P_j(\mathbf{x})} \textrm{d}\mathbf{x} \\
&=&  \frac{ 2^{ \frac{d}{2} } | \mathbf{C}_i |^{\frac{1}{4}} | \mathbf{C}_j |^{\frac{1}{4}} }{ | \mathbf{C}_i + \mathbf{C}_j |^{\frac{1}{2} } } \exp \left[ -\frac{1}{4} ( \boldsymbol{\mu}_i - \boldsymbol{\mu}_j )^T ( \mathbf{C}_i + \mathbf{C}_j)^{-1} ( \boldsymbol{\mu}_i - \boldsymbol{\mu}_j )  \right] \nonumber
\end{eqnarray}
where $\boldsymbol{\mu}$ denotes the position of the Gaussians center and $\mathbf{C}$ its covariance. 

\section{Selecting parameters for reconnaissance metadynamics}

When running a reconnaissance metadynamics trajectory the user must select values for 7 different parameters.  Three of these are the standard metadynamics parameters; namely, the height of the hills, the width of the hills and the frequency of addition.  The remainder deal with the way the adaptive bias is generated by controlling the expansion of basins, establishing how often one should apply the clustering protocol and determining which of the pieces of information obtained from the clustering algorithm should be used in generating CVs.  Guidelines on the setting of each of these parameters is given below.

\subsection{Frequency of storage}

During the trajectory one must periodically collect the values of the collective coordinates as these collected values will be used in the trajectory analysis.  Obviously, it is not sensible to store the collective coordinates at every timestep as the points collected would be very highly correlated.  Instead, one should allow some short period of time to pass prior to collecting statistics as then the extent of correlation will be smaller.  In setting this time one can use the autocorrelation functions to examine what are the typical timescales over which correlations persist.  

\subsection{Frequency of PPCA}

The frequency of PPCA will be determined by the number of points you wish to include in the PPCA analysis.  Setting this parameter is a compromise between the accuracy of the fitting, which will increase with the number of points and hence the time taken to collect the statistics.  In typical applications the time taken for the analysis is insignificant compared with the cost for the dynamics.

\subsection{Weight tolerance}

As discussed in the paper it is only those clusters that do not overlap with the already present basins which have weights greater than some user defined tolerance that are used to create the bias.  Selecting this tolerance is easy, as in considering how to set it one can neglect the effect of overlap.  This means that the tolerance just determines what fraction of the trajectory must be in a basin in order for the user to consider it a significant kinetic trap.

\subsection{Hill height and frequency of addition}

The accuracy of any metadynamics simulations and how it depends on the heights of the hills and the frequency of addition has been studied elsewhere \cite{accuracy_metad, well-tempered}.  Although in reconnaissance metadynamics we are doing something slightly different in terms of hill addition there is no reason to believe that the usual rules do not apply in this case.  Consequentially, so as to not introduce too many discontinuities to the free energy as a function of time we frequently attempt to add hills that have a height lower than $k_B T$ rather than infrequently adding very large hills.

\subsection{Hill width}

The PPCA algorithm provides a measure of the shape of the basin, which the onion-shaped hills are designed to reflect.  As such in calculating the distance $r(\mathbf{s})$ we employ a metric that is equal to the inverse of the covariance matrix.  This ensures that for those directions in which there are large fluctuations the hills are wide, while in directions in which the fluctuations are small the hills are narrow.  This locally adaptive hills shape setting ensures that the bias compensates for the free energy rapidly.  Furthermore, as shown below, it is relatively straightforward to calculate from the covariance matrix the average fluctuations in $r(\mathbf{s})$ so that a sensible width, $\Delta r_i$ for the Gaussians can be selected.   
With no loss of generality, in the following we assume $\boldsymbol{\mu}=0$ and 
$\left<\mathbf{s}\mathbf{s}^T\right>=\mathbf{C}$.
  
\begin{eqnarray}
\sigma[r^2(\mathbf{s}) ] &=& \sqrt{ \left< r(\mathbf{s}) ^4 \right> - \left<r(\mathbf{s}) ^2 \right>^2 } \label{eqn:st-rp2} \\
\left<r(\mathbf{s}) ^2\right> &=& \scriptstyle \left< \mathbf{s}^T \mathbf{C}^{-1} \mathbf{s} \right> \nonumber \\
&=&  \scriptstyle \textrm{Tr} \left< \mathbf{C}^{-1}\mathbf{s}\mathbf{s}^T \right>
= \scriptstyle\textrm{Tr} \left[ \mathbf{C}^{-1} \left<\mathbf{s}\mathbf{s}^T \right> \right]\nonumber  \\
&=& \textrm{Tr} \left[ \mathbf{C}^{-1} \mathbf{C} \right] = d \label{eqn:rp2} \\ 
\left<r(\mathbf{s}) ^4 \right> &=& \scriptstyle \left< \mathbf{s}^T \mathbf{C}^{-1} \mathbf{s} \mathbf{s}^T \mathbf{C}^{-1} \mathbf{s} \right> \nonumber \\
&=& \scriptstyle C_{\alpha \beta} ^{-1} C_{\gamma \delta}^{-1} \left< s_\alpha {s}_\beta {s}_\gamma {s}_\delta \right>  \nonumber \\
&=& \scriptstyle C_{\alpha \beta} ^{-1} C_{\gamma \delta}^{-1} \big[ \left<{s}_\alpha {s}_\beta \right>\left<{s}_\gamma {s}_\delta \right> \nonumber +
\left<{s}_\delta {s}_\alpha \right>\left<{s}_\beta {s}_\gamma \right> \nonumber \\
&+& \scriptstyle \qquad \qquad \quad \left<{s}_\alpha {s}_\gamma \right>\left<{s}_\beta {s}_\delta \right> \big] \label{eqn:identity} \\
&=& \scriptstyle C_{\alpha \beta} ^{-1} C_{\gamma \delta}^{-1} \big[ C_{\alpha \beta}C_{\gamma \delta} +  C_{\delta \alpha}C_{\beta \gamma} + C_{\alpha \gamma}C_{\beta \delta} \big] \nonumber \\
&=& d^2 + d + d = d^2 + 2d \label{eqn:rp4}
\end{eqnarray}    

Here we have used the Einstein summation convention for sums in the later parts and have have used a well-known
equality for the average of products of Gaussian samples\cite{probabily-book}.  
Substituting equations \ref{eqn:rp2} and \ref{eqn:rp4} into equation \ref{eqn:st-rp2} we obtain:
\begin{equation} 
\sigma[r^2(\mathbf{s}) ] = \sqrt{d^2 + 2d - d^2} = \sqrt{2d}     
\end{equation}        

The standard deviation for $r(\mathbf{s})$ can then be obtained using a series expansion as shown below:
\begin{eqnarray}
r(\mathbf{s}) &=& \sqrt{r(\mathbf{s})^2} \qquad  f(r_0^2+\sigma(r^2) ) \sim f(r_0^2) + \sigma(r^2)f'(r_0^2) \nonumber \\
\sigma[r(\mathbf{s}) ] &=& \frac{1}{2\sqrt{\left<r(\mathbf{s})^2\right>}}  \sigma[r(\mathbf{s})^2 ]  = \frac{1}{2\sqrt{d}} \sqrt{2d} = \frac{1}{\sqrt{2}} \label{eqn:st-rp}  
\end{eqnarray}
In metadynamics (and hence reconnaissance metadynamics) the hills should have a width that is a small multiple of this standard deviation.

\subsection{Expansion parameter}

As discussed in the text during reconnaissance metadynamics simulations we must assign a size to the various basins we find from cluster analysis and these sizes change as the simulation progresses.  Obviously, as basins get larger it must become more difficult to grow further as we do not want basins to grow forever.  We therefore use a probabilistic criterion for controlling basin expansion that we derive using some ideas from the theory of diffusion.

The solution of the Fokker Plank equation for diffusion in $d$ dimensions away from the origin is given by:
\begin{equation}
P( \mathbf{s},t ) =  \frac{1}{\sqrt{(2\pi D t)^d}} \exp \left( - \frac{\mathbf{s}^T  \mathbf{s} }{2D t}\right)
\label{eqn:free-diffusion}
\end{equation} 
From this expression we can derive an equation for the average distance between the particle and the origin, $\langle r \rangle$ and the rate of change of this distance:
\begin{equation}
\langle r \rangle(t) = \sqrt{D t} \quad \rightarrow \quad \frac{\textrm{d}\langle r \rangle}{\textrm{d}t} = \frac{1}{2} \sqrt{ \frac{D}{t} } \approx \frac{D}{2\langle r \rangle}
\end{equation}
From this final expression we can obtain an approximate equation for the time taken for the average distance between the particle and the origin to change by a factor of $\Delta r$:  
\begin{equation}
\Delta t' = \frac{2 \Delta r \langle r \rangle}{D}    
\end{equation}
This expression provides us with a measure for the time it would take the average value of the distance from the origin to increase by a factor of $\Delta r$ were the system diffusing about on a flat surface.  This can therefore be used as a speed limit for basin expansion as basins shouldn't expand faster than the system can diffuse away from from the center of the distribution.  In other words, we would like the final basin sizes to reflect the fact that the system was at one time trapped at the outside of a basin and not to reflect the fact that diffusion on a flat free energy surface takes time.   Consequentially, taking the ratio between the time between hill additions, $\Delta t$, and this characteristic time for diffusion provides us with our probabilistic term for basin expansion:
\begin{equation}
P =\textrm{min}\left(1, \frac{D \Delta t}{2 S \Delta r}\right)
\end{equation}
where $S$ is the current size of the basin. 

Obviously, it is not possible to obtain an a-priory estimate for the value of the diffusion constant $D$ in the above expression and thus this parameter must be set by trial and error.  If it is set too small then one will observe that the basins never expand and thus many spurious basins will be observed.  By contrast if it is set too large the basins will never stop expanding. 

\section{Supplementary information: Lennard-Jones 7}

\subsection{Well-tempered metadynamics simulations}

Well tempered metadynamics was used to obtain the free energy surface of the Lennard Jones cluster that is shown in figure 4 in the text.  Four well-tempered metadynamics simulation were run and then averaged so as to increase the speed of convergence of the free energy surface.  In all these simulations $T+\Delta T$, the notional temperature at which the simulation is carried out, was set equal to 1.0~$k_BT$.  The widths of the Gaussians were 0.02 for both CVs and in two of the simulations the heights of the Gaussians were 0.001~$\epsilon$ while in the other two the heights were 0.0025~$\epsilon$.  So as to prevent exploration of uninteresting parts of phase space harmonic walls were used with force constants of 1000~$\epsilon \textrm{CV}^{-2}$.  These were placed at values of 0.4 and 1.2 in the second moment and at a value of -0.3 in the third moment.  

\section{Supplementary information: ala12}

\subsection{Simulating ala12 with unbiassed MD}

Figure S1 shows a representation of a 1~$\mu$s portion of an unbiassed trajectory for alanine 12 modeled with a distance dependent dielectric.  The bars in the lower panel give the values of 40~ps running averages for each of the torsional angles in the protein (see key).  A number of representative structures found during the trajectory are also shown. The system was annealed every 400~ps to examine the transitions between inherent structures and the energies obtained are indicated by the blue line. It is important to note that the timescale in this figure is 20 times as long as the timescale for the reconnaissance metadynamics simulation reported in the paper.  Hence, the figure shows that during a much-longer, unbiassed molecular dynamics simulation the system does not explore a large volume of configuration space and is trapped in a small region of phase space.  In addition this figure shows that the alpha helix is not formed on the timescale shown in the figure. 

\begin{figure*}
\centering
\includegraphics{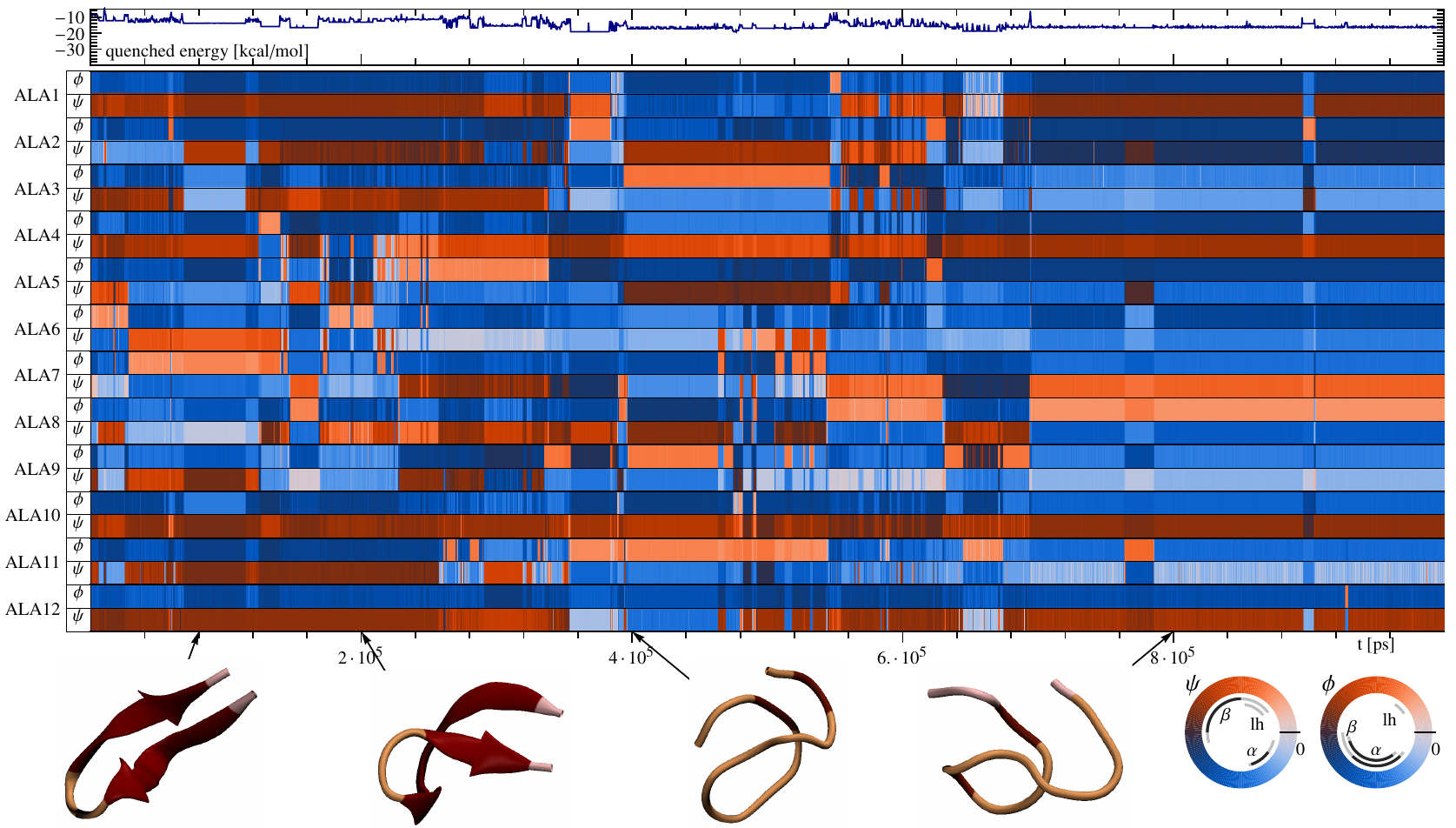}
\caption{A representation of a 1~$\mu$s portion of an unbiassed trajectory for alanine 12 modeled with a distance dependent dielectric.  The bars in the lower panel give the values of 40~ps running averages for each of the torsional angles in the protein (see key).  A number of representative structures found during the trajectory are also shown. The system was annealed every 400~ps to examine the transitions between inherent structures and the energies obtained are indicated by the blue line.}
\label{fig:ala12nobias}
\end{figure*}

\subsection{Quenched structures for ala12}

In addition to this document we also provide a pdb file that contains the quenched structures from a reconnaissance metadynamics trajectory and their energies.  We provide this information as it can serve as a benchmark for the development of new methods for exploring potential energy landscapes.
